\documentclass[aip, jmp, reprint, onecolumn]{revtex4-2}
\usepackage{amsmath}
\usepackage{amssymb}
\usepackage{enumitem}
\usepackage{graphicx}
\usepackage{hyperref}

\allowdisplaybreaks

\begin{document}
\title{Reeh--Schlieder approximation for coherent states}

\author{Riccardo Falcone}
\affiliation{Department of Physics, University of Sapienza, Piazzale Aldo Moro 5, 00185 Rome, Italy}

\author{Claudio Conti}
\affiliation{Department of Physics, University of Sapienza, Piazzale Aldo Moro 5, 00185 Rome, Italy}

\begin{abstract}
We present an explicit, fully local Reeh--Schlieder approximation scheme for coherent states of a free scalar field. For any bounded region $\mathcal{U}$, we construct a one-parameter family of bounded operators $\hat{A}_\zeta$ localized in the causal complement of $\mathcal{U}$. The action of $\hat{A}_\zeta$ on the vacuum approximates the target coherent state in the limit $\zeta \to 0$.
\end{abstract}

\maketitle

\section{Introduction}

The Reeh--Schlieder theorem is a cornerstone result in Algebraic Quantum Field Theory
. It states that for any spacetime region $\mathcal{O}$, the vacuum $|\Omega\rangle$\footnote{For readers with an algebraic background: throughout, we use the GNS representation built from the Minkowski vacuum; $|\Omega\rangle$ denotes the corresponding vacuum vector in that Hilbert space.} is cyclic (and separating) for the associated local algebra $\mathfrak{A}(\mathcal{O})$. In practical terms, acting on the vacuum with operators localized in \textit{any} open spacetime region $\mathcal{O}$ can approximate \textit{any} state in the Hilbert space \footnote{Here the Hilbert space is that of the GNS representation for the Minkowski vacuum.}
arbitrarily well \cite{Reeh1961,haag1992local}. 
Despite its 
breadth, however, the Reeh--Schlieder theorem itself is existential: it guarantees the availability of approximants but does not tell us how to construct them in concrete models, or how to control the approximation error.

In this work, we present an explicit and fully local Reeh--Schlieder approximation scheme for coherent excitations of a free scalar field $\hat{\phi}$. We take $\mathcal{O}$ to be the causal complement of a bounded spacetime region $\mathcal{U}$, i.e., $\mathcal{O} = \mathcal{U}'$. Given a coherent state $|f\rangle=\hat{W}(f)|\Omega\rangle$ generated by a Weyl operator $\hat{W}(f)=\exp [i\hat{\phi}(f)]$ with real test function $f$ and smeared scalar field $\hat{\phi}(f) $, we construct a one-parameter family of bounded operators $\hat{A}_\zeta$ localized in $\mathcal{U}'$ that, when acting on the vacuum $| \Omega \rangle$, approximate the target state $| f \rangle$ to arbitrary accuracy. The associated approximation error
\begin{equation}\label{cyclicity_W}
\mathcal{E}_\zeta(f) = \left\| | f \rangle - \hat{A}_\zeta(f) | \Omega \rangle  \right\|,
\end{equation}
tends to zero as $\zeta \to 0$.

Our construction exploits Tomita--Takesaki modular theory for wedge algebras \cite{takesaki1970tomita,haag1992local} together with the Bisognano--Wichmann 
theorem \cite{10.1063/1.522605,10.1063/1.522898,haag1992local}. For Weyl operators supported in a wedge, we use the analytic continuation of boosts to imaginary rapidity to reflect excitations across the edge of the wedge, and then average along complexified boosts with an analytic mollifier $G_\zeta$ to produce bounded operators supported in $\mathcal{U}'$.

\begin{figure}[h]
\includegraphics[width=0.8\columnwidth]{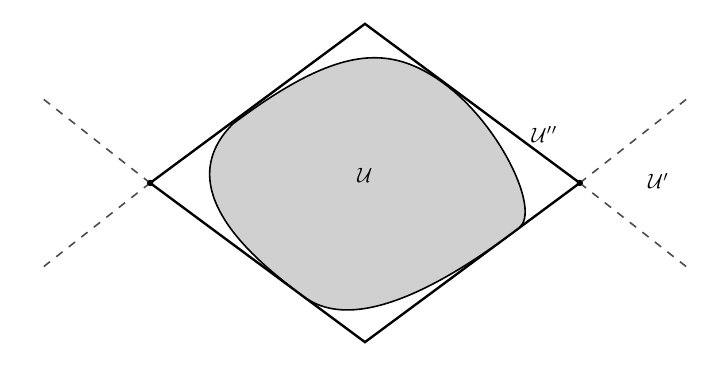}
\caption{Illustration of the causal complement and causal completion of the region $\mathcal{U}$, denoted by $\mathcal{U}'$ and $\mathcal{U}''$, respectively. The causal complement $\mathcal{U}'$ is defined as the set of all points that are spacelike separated from every point in $\mathcal{U}$, whereas $\mathcal{U}''$ is given by the double causal complement, $\mathcal{U}'' = (\mathcal{U}')'$. In the figure, the vertical axis represents the time coordinate $x^0$, and the horizontal axis represents a spatial coordinate.}\label{U}
\end{figure}

The paper is organized as follows. In Sec.~\ref{States_prepared_in_the_causal_region_of_the_detector}, we examine the case where the smearing function $f$ is supported entirely within the causal completion of $\mathcal{U}$ (i.e., its double causal complement), denoted by $\mathcal{U}''$ (see Fig.~\ref{U} for illustration). The opposite situation, where $f$ is supported within $\mathcal{U}'$, is straightforward, since in that case the state is already expressed as $\hat{W}(f) | \Omega \rangle$, with $\hat{W}(f)$ itself a bounded operator localized in $\mathcal{U}'$. In Sec.~\ref{General_unitary_excitation}, we extend the analysis to the more general case where $f$ has arbitrary support, possibly overlapping both $\mathcal{U}'$ and $\mathcal{U}''$. 

\section{Coherent state supported in $\mathcal{U}''$}\label{States_prepared_in_the_causal_region_of_the_detector}

In this section, we consider a coherent state $| f \rangle = \hat{W}(f) | \Omega \rangle$, with $\operatorname{supp}(f) \subseteq \mathcal{U}''$. We construct a one-parameter family of operators $\hat{A}_\zeta(f)$, localized in the causal complement $\mathcal{U}'$, which, when acting on the vacuum $| \Omega \rangle$, approximates the target state $| f \rangle$. The accuracy of this approximation is measured by the error $\mathcal{E}_\zeta(f)$, defined in Eq.~\eqref{cyclicity_W}, which tends to zero as $\zeta \to 0$.

The construction of $\hat{A}_\zeta(f)$ relies on the Tomita--Takesaki modular theory together with the Bisognano--Wichmann theorem. To assist readers who may be unfamiliar with these concepts, and to clarify the notation used throughout this work, we begin with a brief overview in Sec.~\ref{Tomita_Takesaki_operator_and_Bisognano_Wichmann_theorem}. In Sec.~\ref{Weyl_operator_in_the_right_wedge}, we apply this framework to the specific case of a Weyl operator $\hat{W}(f)$ localized in the right wedge, deriving the effect of a Lorentz boost with complex rapidity $\eta + i \pi$ on the coherent state $\hat{W}(f) | \Omega \rangle$. These results are then used in Sec.~\ref{Construction} to construct a family of operators $\hat{A}_\zeta(f) $, each localized in $\mathcal{U}'$, with the property that $\hat{A}_\zeta(f) | \Omega \rangle $ approximates $| f \rangle$ as $\zeta \to 0$. Finally, in Sec.~\ref{Approximation_error}, we provide an explicit expression for the approximation error $\mathcal{E}_\zeta(f) $.

\subsection{Tomita--Takesaki operator and Bisognano--Wichmann theorem}\label{Tomita_Takesaki_operator_and_Bisognano_Wichmann_theorem}

In this subsection, we provide a brief overview of Tomita--Takesaki modular theory and the Bisognano--Wichmann theorem, focusing only on the elements necessary for constructing $\hat{A}_\zeta(f)$: namely, the modular conjugation and modular operator in the right wedge, as well as the analyticity and boundedness properties of Lorentz boosts extended to complex rapidities.

\begin{figure}[h]
\includegraphics[width=0.8\columnwidth]{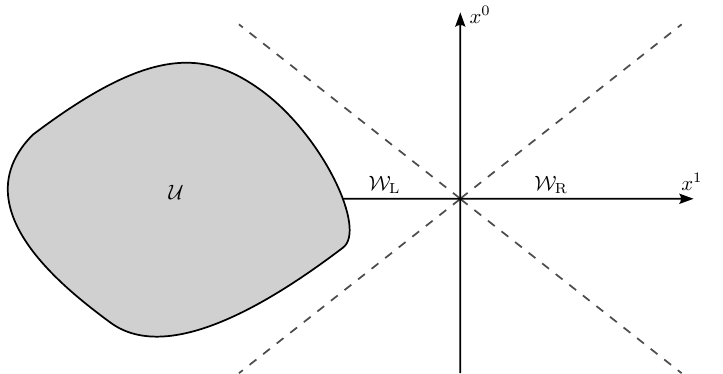}
\caption{The left and right wedges, $\mathcal{W}_\text{L}$ and $\mathcal{W}_\text{R}$, are defined as $\mathcal{W}_\text{L} = \{ x : x^1 < -|x^0| \}$ and $\mathcal{W}_\text{R} = \{ x : x^1 > |x^0| \}$, respectively. The coordinates are chosen such that the region $\mathcal{U}$ lies entirely within $\mathcal{W}_\text{L}$.}\label{wedges}
\end{figure}

Let $\mathcal{W}_\text{L}$ and $\mathcal{W}_\text{R}$ denote the pair of left and right wedges in Minkowski spacetime, identified by $\mathcal{W}_\text{L} = \{ x : x^1<-|x^0| \}$ and $\mathcal{W}_\text{R} = \{ x : x^1>|x^0| \}$, respectively (see Fig.~\ref{wedges}). Consider an operator $\hat{A}_\text{R}$ belonging to the local algebra $\mathfrak{A}(\mathcal{W}_\text{R})$ associated with the right wedge. Within the framework of Tomita--Takesaki theory \cite{takesaki1970tomita, haag1992local}, the Tomita operator $\hat{S}$  is defined by its action on the vector $ \hat{A}_\text{R} | \Omega \rangle$ as
\begin{equation}\label{Tomita}
\hat{S} \hat{A}_\text{R} | \Omega \rangle = \hat{A}_\text{R}^\dagger | \Omega \rangle.
\end{equation}

This operator admits a polar decomposition of the form $\hat{S} = \hat{J} \hat{\Delta}^{1/2}$, where $\hat{J}$ is an anti-unitary operator known as the modular conjugation, and $\hat{\Delta}$ is a positive, self-adjoint operator called the modular operator. The explicit characterization of $\hat{J}$ and $\hat{\Delta}$ is provided by the Bisognano--Wichmann theorem \cite{10.1063/1.522605, 10.1063/1.522898, haag1992local}. Specifically, $\hat{J}$ implements a CPT transformation combined with a spatial rotation by an angle $\pi$ about the $x^1$-axis. The action of $\hat{J}$ on the smeared field is given by
\begin{equation}\label{J_phi_J}
\hat{J} \hat{\phi}(f) \hat{J} =  \hat{\phi}(f \circ J),
\end{equation}
where $J$ denotes the spacetime reflection in the $x^0$ and $x^1$ directions, explicitly defined as
\begin{equation}
J \! \left(x^0, x^1, x^2, x^3\right) = \left(-x^0, -x^1, x^2, x^3\right).
\end{equation}

The modular operator $\hat{\Delta}$, on the other hand, is related to the Lorentz boost transformations in the $x^1$-direction, which we denote by $\Lambda_1(\eta)$. These boosts act on a spacetime point $x = (x^0, x^1, x^2, x^3)$ as
\begin{equation}\label{Lambda_1_eta}
\Lambda_1(\eta)\!\left(x^0, x^1, x^2, x^3\right) = \left[\cosh (\eta) x^0 + \sinh (\eta) x^1, \cosh (\eta) x^1 + \sinh (\eta) x^0, x^2, x^3\right].
\end{equation}
The unitary implementation of $\Lambda_1(\eta)$, written as $\hat{U}[\Lambda_1(\eta)]$, acts on the smeared field operator as
\begin{equation}\label{U_Lambda_phi_f_R}
\hat{U}[\Lambda_1(\eta)] \hat{\phi}(f) \{ \hat{U}[\Lambda_1(\eta)] \}^\dagger = \hat{\phi}\!\left[f \circ \Lambda_1(-\eta)\right].
\end{equation}
This relation can be understood as first applying the pullback of $\Lambda_1(\eta)$ to the unsmeared field $\hat{\phi}$, followed by integration by parts against the test function $f$.

The square root of the modular operator $\hat{\Delta}^{1/2}$ is realized by the unitary representation of the boost $\Lambda_1(\eta)$ evaluated at imaginary rapidity $\eta = i \pi$, i.e.,
\begin{equation}
\hat{\Delta}^{1/2} = \hat{U}\!\left[\Lambda_1(i \pi)\right].
\end{equation}
Explicitly, this unitary operator can be written as $\hat{U}[\Lambda_1(i \pi)] = \exp(- \pi \hat{K}_1)$, where $\hat{K}_1$ is the boost generator defined through $\hat{U}[\Lambda_1(\eta)] = \exp(i \eta \hat{K}_1)$.

By applying $\hat{J}$ to both sides of Eq.~\eqref{Tomita}, and using the identity $\hat{J}^2 = \hat{\mathbb{I}}$ and the invariance of the vacuum under modular conjugation, $ \hat{J}| \Omega \rangle = | \Omega \rangle$, we obtain
\begin{equation}\label{Tomita_2}
\hat{U}\!\left[\Lambda_1(i \pi)\right] \hat{A}_\text{R} | \Omega \rangle = \hat{J} \hat{A}_\text{R}^\dagger \hat{J} | \Omega \rangle,
\end{equation}
where the operator $ \hat{J} \hat{A}_\text{R}^\dagger \hat{J}$ on the right-hand side belongs to the local algebra $\mathfrak{A}(\mathcal{W}_\text{L})$, meaning it is localized in the causal complement of $\mathcal{W}_\text{R}$. In what follows, we will make use of Eq.~\eqref{Tomita_2}, along with the fact that the map $\eta \mapsto \hat{U}[\Lambda_1(\eta)] \hat{A}_\text{R} | \Omega \rangle$ is analytic within the open strip $0<\Im(\eta)<\pi$, and strongly continuous and bounded on the closed strip $0 \leq \Im(\eta) \leq \pi$ \cite{10.1063/1.522605, 10.1063/1.522898, haag1992local}.

\subsection{Weyl operator in the right wedge}\label{Weyl_operator_in_the_right_wedge}

In Sec.~\ref{Tomita_Takesaki_operator_and_Bisognano_Wichmann_theorem}, we derived Eq.~\eqref{Tomita_2} by using the Bisognano--Wichmann theorem. We now apply this result to the specific case where the operator $\hat{A}_\text{R}$ in Eq.~\eqref{Tomita_2} is taken to be a Weyl operator $\hat{W}(f_\text{R})$. To ensure that $\hat{W}(f_\text{R})$ belongs to the algebra associated with the right wedge, we require that the test function $f_\text{R}$ is supported entirely within that region. In this case, Eq.~\eqref{Tomita_2} takes the form
\begin{equation}\label{Tomita_3}
\hat{U}\!\left[\Lambda_1(i \pi)\right] \hat{W}(f_\text{R}) | \Omega \rangle = \hat{J} \hat{W}(-f_\text{R}) \hat{J} | \Omega \rangle.
\end{equation}

By invoking Eq.~\eqref{J_phi_J} and the antilinearity of $\hat{J}$, this expression can be rewritten as
\begin{equation}\label{Tomita_4}
\hat{U}\!\left[\Lambda_1(i \pi)\right] \hat{W}(f_\text{R}) | \Omega \rangle = \hat{W}(f_\text{R} \circ J) | \Omega \rangle.
\end{equation}

Equation \eqref{Tomita_4} describes the effect of a Lorentz boost with rapidity $i \pi$  on the coherent state $\hat{W}(f_\text{R}) | \Omega \rangle$. This result can be generalized to Lorentz boosts with complex rapidity $\eta + i \pi$, by applying $\hat{U}[\Lambda_1(\eta)]$ to both sides. By using Eq.~\eqref{U_Lambda_phi_f_R}, along with the invariance of the vacuum under Lorentz boosts, $\{ \hat{U}[\Lambda_1(\eta)] \}^\dagger | \Omega \rangle = | \Omega \rangle$, we obtain the relation $\hat{U}[\Lambda_1(\eta)] \hat{W}(f_\text{R} \circ J) | \Omega \rangle =  \hat{W}[f_\text{R} \circ J \circ \Lambda_1(-\eta) ] | \Omega \rangle$. This identity allows us to compute the action of $\hat{U}[\Lambda_1(\eta)]$ on the right-hand side of Eq.~\eqref{Tomita_4}. On the left-hand side, we invoke the group composition rule for Lorentz boosts: $\hat{U}[\Lambda_1(\eta)] \hat{U}[\Lambda_1(i \pi)] = \exp(i \eta \hat{K}_1) \exp( - \pi \hat{K}_1)  = \exp[i (\eta + i\pi) \hat{K}_1 ] =  \hat{U}[\Lambda_1(\eta + i \pi)]$. By combining these observations, we arrive at
\begin{equation}\label{Tomita_5}
\hat{U}\!\left[\Lambda_1(\eta + i \pi)\right] \hat{W}(f_\text{R}) | \Omega \rangle = \hat{W}\!\left[f_\text{R} \circ J \circ \Lambda_1(-\eta)\right] | \Omega \rangle,
\end{equation}
which holds for all $\eta \in \mathbb{R}$.

\subsection{Construction of $\hat{A}_\zeta(f) $}\label{Construction}

In Sec.~\ref{Tomita_Takesaki_operator_and_Bisognano_Wichmann_theorem}, we introduced the left and right wedges and presented the Bisognano--Wichmann theorem for operators $\hat{A}_\text{R}$ localized in the right wedge. We also noted the analyticity of $\hat{U}[\Lambda_1(\eta)] \hat{A}_\text{R} | \Omega \rangle$ for $0<\Im(\eta)<\pi$, as well as its continuity and boundedness on the closed strip $0 \leq \Im(\eta) \leq \pi$. Then, in Sec.~\ref{Weyl_operator_in_the_right_wedge}, we focused on Weyl operators $\hat{W}(f_\text{R})$ localized in the right wedge and derived the action of Lorentz boosts with complex rapidity $\eta + i \pi$  on the corresponding coherent state $\hat{W}(f_\text{R}) | \Omega \rangle$. In this subsection, we return to our original objective. Given a coherent state $| f \rangle = \hat{W}(f) | \Omega \rangle$, with the test function $f$ fully supported within the region $\mathcal{U}''$, we use the previous results to construct a one-parameter family of bounded operators $\hat{A}_\zeta(f) $, each localized in $\mathcal{U}'$. As part of the construction, we require that the approximation error $\mathcal{E}_\zeta(f) $, defined in Eq.~\eqref{cyclicity_W}, vanishes in the limit $\zeta \to 0$.

To relate this construction to the framework of Tomita--Takesaki modular theory and the Bisognano--Wichmann theorem, we choose a Minkowski coordinate system in which the region $\mathcal{U}$ lies entirely within the left wedge $\mathcal{W}_\text{L}$ (see Fig.~\ref{wedges}). In this configuration, its causal completion $\mathcal{U}''$ is also contained within $\mathcal{W}_\text{L}$. Since the test function $f$ is supported in $\mathcal{U}'' \subset \mathcal{W}_\text{L}$, the reflected function $f \circ J$ is supported entirely in the right wedge $\mathcal{W}_\text{R}$ and can thus serve as the function $f_\text{R}$ of Eq.~\eqref{Tomita_5}. This leads to the identity
\begin{equation}\label{Tomita_6}
\hat{U}\!\left[\Lambda_1(\eta + i \pi)\right] \hat{W}(f \circ J) | \Omega \rangle = \hat{W}\!\left[f \circ \Lambda_1(-\eta)\right] | \Omega \rangle,
\end{equation}
which holds because $J \circ J = \mathbb{I}$.

Based on this result, we define
\begin{equation}\label{W_prime_zeta}
\hat{A}_\zeta(f) = \int_\mathbb{R} d\eta \, G_\zeta(\eta - i \pi) \, \hat{W}\!\left[f \circ J \circ \Lambda_1(-\eta)\right],
\end{equation}
where $G_\zeta(\eta)$ is a distribution that satisfies the following conditions:
\begin{enumerate}[label=(\roman*)]
\item it is analytic within the strip $-\pi < \Im(\eta) < 0$;\label{G_zeta_analytic}
\item it decays for large $|\Re(\eta)|$ within this strip, i.e., $  G_\zeta(\eta) \to 0$ as $|\Re(\eta)| \to \infty$ for $\Im(\eta) \in (-\pi,0)$; \label{G_zeta_decay}
\item in the distributional limit $\zeta \to 0$, it converges to the Dirac delta function $\delta(\eta)$ on the real axis. \label{G_zeta_delta}
\end{enumerate}

Since the operator $\hat{W}[f \circ J \circ \Lambda_1(-\eta)]$ belongs to the local algebra of the right wedge, it follows that $\hat{A}_\zeta(f)$ is localized within the right wedge. As a result, $\hat{A}_\zeta(f)$ is also localized in the causal complement of $\mathcal{U}$, thereby satisfying one of the key requirements for this family of operators.

To compute the action of $\hat{A}_\zeta(f)$ on the vacuum, we start by using Eqs.~\eqref{U_Lambda_phi_f_R} and \eqref{W_prime_zeta}, along with the invariance of the vacuum under Lorentz boosts, $\{ \hat{U}[\Lambda_1(\eta)] \}^\dagger | \Omega \rangle = | \Omega \rangle$. This yields
\begin{equation}\label{W_prime_zeta_Omega}
\hat{A}_\zeta(f) | \Omega \rangle = \int_\mathbb{R} d\eta \, G_\zeta(\eta - i \pi) \, \hat{U}\!\left[\Lambda_1(\eta)\right] \, \hat{W}(f \circ J ) | \Omega \rangle.
\end{equation}

Next, we invoke Cauchy's theorem, 
choosing the boundary of the strip $0 < \Im(\eta) < \pi$ as the integration contour and taking the vector-valued function $ G_\zeta(\eta - i \pi) \hat{U}[\Lambda_1(\eta)] \hat{W}(f \circ J) | \Omega \rangle$ as the integrand. By using the analyticity of $G_\zeta(\eta)$ [property \ref{G_zeta_analytic}], the decay condition on $G_\zeta(\eta)$ along the lateral edges of the contour [property \ref{G_zeta_decay}] and the analyticity and boundedness of the map $\eta \mapsto \hat{U}[\Lambda_1(\eta)] \hat{W}(f \circ J) | \Omega \rangle$  \cite{10.1063/1.522605, 10.1063/1.522898, haag1992local}, we obtain the identity
\begin{equation}
\int_\mathbb{R} d\eta \, G_\zeta(\eta - i \pi) \, \hat{U}\!\left[\Lambda_1(\eta)\right] \hat{W}(f \circ J) | \Omega \rangle = \int_\mathbb{R} d\eta \, G_\zeta(\eta) \, \hat{U}\!\left[\Lambda_1(\eta + i \pi)\right] \hat{W}(f \circ J) | \Omega \rangle.
\end{equation}
By combining this result with Eqs.~\eqref{Tomita_6} and \eqref{W_prime_zeta_Omega}, we find that the action of $\hat{A}_\zeta(f)$ on the vacuum $| \Omega \rangle$ is given by
\begin{equation}\label{W_prime_zeta_f_Omega}
\hat{A}_\zeta(f) | \Omega \rangle = \int_\mathbb{R} d\eta \, G_\zeta(\eta) \, \hat{W}\!\left[f \circ \Lambda_1(-\eta)\right] | \Omega \rangle.
\end{equation}

Due to property \ref{G_zeta_delta}, the right hand side of Eq.~\eqref{W_prime_zeta_f_Omega} converges to $\hat{W}[f \circ \Lambda_1(0)] | \Omega \rangle = \hat{W}( f) | \Omega \rangle = | f \rangle$ as $\zeta \to 0$. This means that the operator $\hat{A}_\zeta(f)$ approximates the coherent state $| f \rangle$ in the limit of small $\zeta$ by acting on the vacuum $| \Omega \rangle$.

\subsection{Approximation error $\mathcal{E}_\zeta(f) $}\label{Approximation_error}

In Sec.~\ref{Construction}, we introduced a one-parameter family of operators $\hat{A}_\zeta(f) $, each localized in $\mathcal{U}'$, constructed so that $\hat{A}_\zeta(f) | \Omega \rangle $ approximates $| f \rangle$ as $\zeta \to 0$. In this subsection, we derive an explicit expression for the approximation error $\mathcal{E}_\zeta(f) $, as defined in Eq.~\eqref{cyclicity_W}
.

To obtain a concrete expression for $\mathcal{E}_\zeta(f) $, we must specify a form for the distribution $G_\zeta(\eta)$. A natural and convenient choice that satisfies conditions \ref{G_zeta_analytic}–\ref{G_zeta_delta} is the Gaussian function
\begin{equation}\label{G_zeta_gaussian}
G_\zeta(\eta) = \frac{1}{\sqrt{2 \pi \zeta}} \exp \! \left( - \frac{\eta^2}{2 \zeta}  \right).
\end{equation}

By using Eqs.~\eqref{cyclicity_W} and \eqref{W_prime_zeta_f_Omega}, the approximation error $\mathcal{E}_\zeta(f)$ can be expressed as
\begin{equation}
\mathcal{E}_\zeta(f) = \left\| \int_\mathbb{R} d\eta \,[\delta(\eta) - G_\zeta(\eta)] \left|f \circ \Lambda_1(-\eta)  \right\rangle  \right\|,
\end{equation}
or, more explicitly,
\begin{equation}\label{Reeh_Schlieder_approximation_error}
\mathcal{E}_\zeta(f) = \sqrt{ \int_\mathbb{R} d\eta \, [\delta(\eta) - G_\zeta(\eta)]  \int_\mathbb{R} d\eta' \, [\delta(\eta') - G_\zeta(\eta')] \left\langle  f \circ \Lambda_1(-\eta) \middle|f \circ \Lambda_1(-\eta')  \right\rangle } .
\end{equation}

The inner product between two arbitrary coherent states $| f_1 \rangle$ and $| f_2 \rangle$ is
\begin{equation}
\langle f_1 | f_2 \rangle = \exp \! \left[ W_2(f_1, f_2) - \frac{1}{2} W_2(f_1, f_1)  - \frac{1}{2} W_2(f_2, f_2) \right],
\end{equation}
where
\begin{equation}\label{W_2}
W_2(f_1,f_2) = \left\langle \Omega \middle| \hat{\phi}(f_1) \hat{\phi}(f_2) \middle| \Omega \right\rangle
\end{equation}
is the Wightman two-point function smeared with test functions $f_1$ and $f_2$. By setting $f_1 = f \circ \Lambda_1(-\eta)$ and $f_2 = f \circ \Lambda_1(-\eta') $, we obtain the product between two coherent states of the form $\left|f \circ \Lambda_1(-\eta)  \right\rangle $ as
\begin{align}\label{Lambda_f_product_eta_eta_prime}
& \left\langle f \circ \Lambda_1(-\eta) \middle|f \circ \Lambda_1(-\eta')  \right\rangle
= \exp \left\lbrace  W_2 \! \left[  f \circ \Lambda_1(-\eta),  f \circ \Lambda_1(-\eta') \right] \right. \nonumber \\
& \left. - \frac{1}{2} W_2 \! \left[  f \circ \Lambda_1(-\eta),  f \circ \Lambda_1(-\eta) \right]  - \frac{1}{2} W_2 \! \left[  f \circ \Lambda_1(-\eta'),  f \circ \Lambda_1(-\eta') \right]  \right\rbrace.
\end{align}

By using Eq.~\eqref{U_Lambda_phi_f_R}, the invariance of the vacuum under Lorentz boosts, $\{ \hat{U}[\Lambda_1(\eta)] \}^\dagger | \Omega \rangle = | \Omega \rangle$, and the composition property of boosts, $\{\hat{U}[\Lambda_1(\eta)]\}^\dagger \hat{U}[\Lambda_1(\eta')] =  \hat{U}[\Lambda_1(\eta' - \eta)]$, we obtain
\begin{equation}
W_2\!\left[ f \circ \Lambda_1(-\eta), f \circ \Lambda_1(-\eta') \right] = W_2\!\left[ f, f \circ \Lambda_1(\eta - \eta') \right].
\end{equation}
Hence, Eq.~\eqref{Lambda_f_product_eta_eta_prime} simplifies to
\begin{equation}\label{Lambda_f_product_eta_eta_prime_2}
\left\langle f \circ \Lambda_1(-\eta) \middle|f \circ \Lambda_1(-\eta')  \right\rangle = \exp \! \left\lbrace W_2 \! \left[ f,  f \circ \Lambda_1(\eta - \eta') \right] - W_2 (f,f) \right\rbrace.
\end{equation}
This result allows us to rewrite Eq.~\eqref{Reeh_Schlieder_approximation_error} as
\begin{align}\label{Reeh_Schlieder_approximation_error_2}
& \mathcal{E}_\zeta(f)\nonumber \\
 = & \sqrt{ \int_\mathbb{R} d\eta \, [\delta(\eta) - G_\zeta(\eta)] \int_\mathbb{R} d\eta' \, [\delta(\eta') - G_\zeta(\eta')] \, \exp \left\lbrace W_2 \! \left[ f,  f \circ \Lambda_1(\eta - \eta') \right] - W_2 (f,f)  \right\rbrace } .
\end{align}

We now employ the Gaussian convolution identity
\begin{equation}\label{Gaussian_convolution}
\int_\mathbb{R} d\eta \, G_\zeta(\eta) \int_\mathbb{R} d\eta' \,  G_\zeta(\eta') \,  g(\eta - \eta')   = \int_\mathbb{R} d\eta \, G_{2 \zeta}(\eta) g(\eta),
\end{equation}
which holds for any function $g(\eta)$
. By applying the identity \eqref{Gaussian_convolution} to Eq.~\eqref{Reeh_Schlieder_approximation_error_2} and performing a change of variables $\eta' \mapsto - \eta'$ where needed, we finally obtain
\begin{equation}\label{epsilon_zeta_F}
\mathcal{E}_\zeta(f) = \sqrt{1 - \int_\mathbb{R} d\eta  [ 2 G_\zeta(\eta) -  G_{2 \zeta}(\eta)]  \exp \left\lbrace W_2 \! \left[ f,  f \circ \Lambda_1(\eta) \right] - W_2 (f,f) \right\rbrace } .
\end{equation}
To verify that $\mathcal{E}_\zeta(f) \to 0$ as  $\zeta \to 0$, it suffices to note that the Gaussian distribution $G_\zeta(\eta) $ converges to the Dirac delta function $\delta(\eta)$ in this limit, and that $W_2 [ f,  f \circ \Lambda_1(0) ] = W_2 (f,f)$.

\section{General coherent state}\label{General_unitary_excitation}

In this section, we consider a coherent state of the form $| f \rangle = \hat{W}(f) | \Omega \rangle$, where the smearing function $f$ is not subject to any particular constraint. The approximation method introduced in Sec.~\ref{States_prepared_in_the_causal_region_of_the_detector} remains applicable here, provided the coordinate system is chosen so that the left wedge $\mathcal{W}_\text{L}$ encloses the enlarged region $ \mathcal{U} \cup \operatorname{supp}(f)$, rather than just $\mathcal{U}$. This construction, however, ceases to be applicable in the particular case where the set of all points causally connected to $\operatorname{supp}(f)$ coincides with the whole of Minkowski spacetime.

\begin{figure}[h]
\includegraphics[width=0.8\columnwidth]{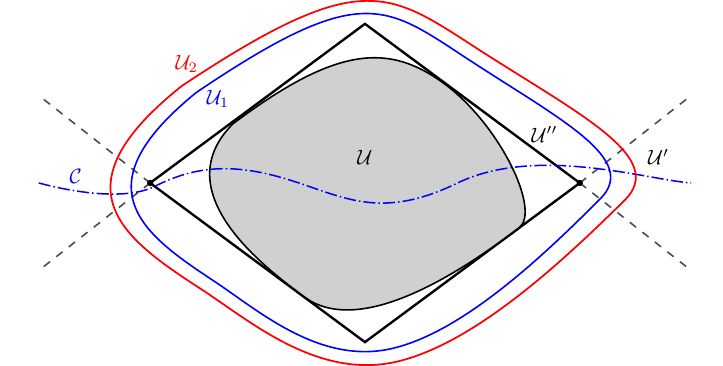}
\caption{The regions $\mathcal{U}'$ and $\mathcal{U}''$ denote the causal complement and causal completion of $\mathcal{U}$, respectively. We then introduce two nested extensions, $\mathcal{U}_1$ and $\mathcal{U}_2$, chosen so that $\mathcal{U}'' \subset \mathcal{U}_1 \subset \mathcal{U}_2$. The union $\mathcal{U}_1 \cup \mathcal{U}'$ contains a Cauchy surface $\mathcal{C}$. In the figure, the vertical axis corresponds to the time coordinate $x^0$, while the horizontal axis represents a spatial coordinate.}\label{U_1_2}
\end{figure}

To address this issue, let us enlarge the region $\mathcal{U}''$ to $\mathcal{U}_1  \supset \mathcal{U}''$ (see Fig.~\ref{U_1_2}). For any such extension, the union $\mathcal{U}_1 \cup \mathcal{U}'$ always contains a Cauchy surface for the entire spacetime. The presence of a Cauchy surface in this union ensures the existence of a function $f_0$ supported entirely within $\mathcal{U}_1 \cup \mathcal{U}'$, such that $\hat{W}(f) = \hat{W}(f_0)$. This property is known as the time-slice property of the field \cite{Dimock1980}. Physically, it reflects the fact that, due to the dynamical laws, any field operator can be expressed in terms of field data on an arbitrarily small neighborhood of a Cauchy surface.

Next, consider a further enlargement $\mathcal{U}_2 \supset \mathcal{U}_1$ (see Fig.~\ref{U_1_2}) and choose a smooth bump function $\chi$ satisfying $0\leq\chi\leq 1$, $\chi |_{\mathcal{U}_1} = 1$ and $\operatorname{supp}(\chi) = \mathcal{U}_2$. This function provides a smooth localization of any test function to $\mathcal{U}_1$, using the collar region $\mathcal{U}_2 \setminus \mathcal{U}_1$ as a transition zone. By introducing the complementary cutoff $\chi' = 1 - \chi$ and recalling that $\operatorname{supp}(f_0) \subseteq \mathcal{U}_1 \cup \mathcal{U}'$, we obtain the decomposition $f_0 = \chi f_0+ \chi' f_0$, where $\chi f_0$ is supported in $\mathcal{U}_2 \cap (\mathcal{U}_1 \cup \mathcal{U}')$ and $\chi' f_0$ is supported in $\mathcal{U}' \setminus \mathcal{U}_1$.

By using the time-slice identity $\hat{W}(f) = \hat{W}(f_0)$, the decomposition $f_0 = \chi f_0+ \chi' f_0$ and the Weyl-form canonical commutation relation 
$\hat{W}(f_1)\hat{W}(f_2) = \exp\{ - i \Im[ W_2(f_1,f_2) ] \} \hat{W}(f_1+f_2)$,
we find that the Weyl operator $\hat{W}(f)$ factorizes as
\begin{equation}\label{W_canonical_2}
\hat{W}(f) = \exp\!\left\lbrace i \Im\!\left[ W_2\!\left(\chi' f_0,\chi f_0\right) \right] \right\rbrace \hat{W}(\chi' f_0) \hat{W}(\chi f_0).
\end{equation}
This expresses $\hat{W}(f)$ as the product of a phase factor $\exp \{ i \Im [ W_2(\chi' f_0,\chi f_0) ] \}$ and two unitary operators: $ \hat{W}(\chi' f_0)$, localized in $\mathcal{U}' \setminus \mathcal{U}_1 \subset \mathcal{U}' $, and $\hat{W}(\chi f_0)$, localized in $\mathcal{U}_2 \cap (\mathcal{U}_1 \cup \mathcal{U}') \subset \mathcal{U}_2$.

In choosing the spacetime coordinates $(x^0, x^1, x^2, x^3)$, we now fix them so that the left wedge $\mathcal{W}_\text{L} $ contains the extended region $\mathcal{U}_2$, rather than $\mathcal{U}$, as in Sec.~\ref{States_prepared_in_the_causal_region_of_the_detector}. Once these adjustments are implemented, we can proceed using the same method outlined in Sec.~\ref{States_prepared_in_the_causal_region_of_the_detector} to approximate the state $\hat{W}(\chi f_0) | \Omega \rangle$ by means of the family of vectors $\int_\mathbb{R} d\eta G_\zeta(\eta - i \pi)  \hat{W}[\chi f_0\circ J \circ \Lambda_1(-\eta)] | \Omega \rangle $ parametrized by the variable $\zeta$.

As a result, we find that the family of operators
\begin{equation}
\hat{A}_\zeta(f) = \exp\!\left\lbrace i \Im\!\left[ W_2\!\left(\chi' f_0,\chi f_0\right) \right] \right\rbrace  \hat{W}(\chi' f_0)  \int_\mathbb{R} d\eta \, G_\zeta(\eta - i \pi) \, \hat{W}\!\left[\chi f_0\circ J \circ \Lambda_1(-\eta)\right],
\end{equation}
localized in $\mathcal{U}'$, approximate the target state $| f \rangle$ by acting on the vacuum $| \Omega \rangle$. The approximation error is
\begin{equation}\label{epsilon_zeta_F_2}
\mathcal{E}_\zeta(f) = \sqrt{1 - \int_\mathbb{R} d\eta  [ 2 G_\zeta(\eta) -  G_{2 \zeta}(\eta)]  \exp \left\lbrace W_2 \! \left[ \chi f_0,  \chi f_0\circ \Lambda_1(\eta) \right] - W_2 (\chi f_0,\chi f_0) \right\rbrace },
\end{equation} 
which vanishes as $\zeta \to 0$. The validity of Eq.~\eqref{epsilon_zeta_F_2} follows from Eqs.~\eqref{cyclicity_W} and \eqref{W_canonical_2} and the computation in Sec.~\ref{Approximation_error}, together with the observation that the phase factor $\exp \{ i \Im [ W_2(\chi' f_0,\chi f_0) ] \}$ and the unitary operator $\hat{W}(\chi' f_0)$ leave the norm of any state invariant. 


\section*{Acknowledgment}

We acknowledge support from HORIZON EIC-2022-PATHFINDERCHALLENGES-01 HEISINGBERG Project No.~101114978.

\bibliography{bibliography}

\end{document}